\documentclass[aps,twocolumn,showpacs,amssymb,prb,superscriptaddress]{revtex4-1}

\usepackage{amsmath}
\usepackage{hyperref}
\usepackage{graphicx}
\usepackage{bbm}
\usepackage{color}

\usepackage{bm}

\allowdisplaybreaks

\begin{document}
\title{Spin transport through a spin-1/2 XXZ chain contacted to fermionic leads}

\author{Florian Lange}
\affiliation{Institut f\"ur Physik, Ernst-Moritz-Arndt-Universit\"at
Greifswald, 17489 Greifswald, Germany}
\affiliation{Computational Condensed Matter Physics Laboratory, RIKEN Cluster for Pioneering Research (CPR), Saitama 351-0198, Japan}

\author{Satoshi Ejima}
\affiliation{Institut f\"ur Physik, Ernst-Moritz-Arndt-Universit\"at
Greifswald, 17489 Greifswald, Germany}
\affiliation{Computational Condensed Matter Physics Laboratory, RIKEN Cluster for Pioneering Research (CPR), Saitama 351-0198, Japan}

\author{Tomonori Shirakawa}
\affiliation{International School for Advanced Studies (SISSA), via Bonomea 265, 34136, Trieste, Italy}
\affiliation{Computational Condensed Matter Physics Laboratory, RIKEN Cluster for Pioneering Research (CPR), Saitama 351-0198, Japan}
\affiliation{Computational Quantum Matter Research Team, RIKEN Center for Emergent Matter Science (CEMS), Saitama 351-0198, Japan}
\affiliation{Computational Materials Science Research Team, RIKEN Center for Computational Science (R-CCS),  Hyogo 650-0047,  Japan}

\author{Seiji Yunoki}
\affiliation{Computational Condensed Matter Physics Laboratory, RIKEN Cluster for Pioneering Research (CPR), Saitama 351-0198, Japan}
\affiliation{Computational Quantum Matter Research Team, RIKEN Center for Emergent Matter Science (CEMS), Saitama 351-0198, Japan}
\affiliation{Computational Materials Science Research Team, RIKEN Center for Computational Science (R-CCS),  Hyogo 650-0047,  Japan}

\author{Holger Fehske}
\affiliation{Institut f\"ur Physik, Ernst-Moritz-Arndt-Universit\"at
Greifswald, 17489 Greifswald, Germany}

\begin{abstract}
We employ matrix-product state techniques to numerically study the zero-temperature spin transport in 
a finite spin-1/2 XXZ chain coupled to fermionic leads with a spin bias voltage. 
Current-voltage characteristics are calculated for parameters corresponding to the gapless XY phase and the 
gapped N\'eel phase. In both cases, the low-bias spin current is strongly suppressed unless the parameters of 
the model are fine-tuned. For the XY phase, this corresponds to a conducting fixed point  
where the conductance agrees with the Luttinger-liquid prediction. 
In the N\'eel phase, fine-tuning the parameters similarly leads to an unsuppressed spin current with a linear 
current-voltage characteristic at low bias voltages. 
However, with increasing the bias voltage, 
there occurs a sharp crossover to a region where the current-voltage characteristic is no longer linear and 
a smaller differential conductance is observed.  
We furthermore show that the parameters maximizing the spin current minimize the Friedel oscillations at the interface, 
in agreement with the previous analyses of the charge current for inhomogeneous Hubbard and spinless fermion chains. 
\end{abstract}

\date{\today}

\maketitle

\section{Introduction}

Besides the more usual semiconductor- and metal-based spintronics, 
there have been proposals to use magnetic insulators in spin-based devices~\cite{SpinWaveSpinCurrent,SpinSeebeckInsulators,SpinonSpinCurrent}. 
An advantage of these systems would be the absence of scattering due to conduction electrons, which may allow 
spin-current transmission over longer distances. 
Experiments have demonstrated the possibility to electrically induce a magnon spin current at a Pt/Y$_3$Fe$_5$O$_{12}$ interface by using the spin-Hall effect~\cite{SpinWaveSpinCurrent}. 
More recently, a spin current has been driven through the spin-1/2-chain material Sr$_2$CuO$_3$ 
by applying a temperature gradient~\cite{SpinonSpinCurrent}. 
This was interpreted as a spinon spin current induced by the spin-Seebeck effect.

A lot of research has been reported on the spin transport in the antiferromagnetic spin-1/2 XXZ chain, 
especially concerning the question whether the dynamics are ballistic or diffusive in the linear-response regime. 
At zero temperature, it is known from the exact Bethe-ansatz calculations that the spin transport 
is ballistic in the gapless phase and diffusive in the gapped phase~\cite{ShastrySutherland}. 
There is considerable analytical and numerical evidence that this also holds true 
at any finite temperature~\cite{Zotos,XXZMeisner2003,Znidaric,XXZKarrasch2013,Prosen2011}. 
A possible exception is the $SU(2)$ isotropic point for which differing results have been obtained.

Here, we study the finite-bias spin transport for a specific setup with 
fermionic leads at zero temperature.
To this end, we employ the density-matrix renormalization group (DMRG)~\cite{White92} and the real-time evolution of matrix-product states (MPS) via the time-evolving block decimation (TEBD)~\cite{PhysRevLett.91.147902}.  
The difference from previous studies of transport in finite spin chains is our choice of the leads. 
In Refs.~\onlinecite{Prosen2009,PhysRevB.80.035110,NDCSpinChain}, boundary driving modeled by a Lindblad equation was considered, which allows the direct calculation of the non-equilibrium steady state with matrix-product-operator techniques. Interestingly, a negative differential conductance was observed for strong driving in the gapped phase. 
Other studies have explored the transport in inhomogeneous XXZ chains~\cite{SpinCurrentRectification} and 
fermionic quantum wires coupled to non-interacting leads, which map to an XXZ chain through a Jordan-Wigner transformation~\cite{SchmitteckertTransport2004,DMRGKubo,ReviewSchmitteckert,ConductingFixedPoint1,Ponomarenko1999}.

In setups with leads, the transport may 
be influenced by backscattering at the interfaces which, for repulsive interactions, can completely inhibit transport 
at low voltages and temperatures~\cite{KaneFisher1,KaneFisher2}. 
In general, the strength of the backscattering 
will depend in a non-trivial way on the parameters on either side of the interface. 
In particular, it has been shown for typical models of fermionic chains that conducting fixed points with perfect 
conductance exist\cite{ConductingFixedPoint1,ConductingFixedPoint2,ConductingFixedPoint3}.

The primary concern of this paper is to numerically explore the possibility of such conducting fixed points for our specific 
setup of the junction. 
We consider both the gapless XY and the gapped N\'eel phase of the spin-1/2 XXZ chain. 
In the latter case, the energy gap leads to insulating behavior at zero temperature.  
One may then ask, how the insulating state breaks down at finite bias voltage 
and how the transport depends on the length of the chain. 
The charge transport in a similar setup with a Mott-insulating Hubbard chain 
has been addressed, e.g., in Ref.~\onlinecite{MeisnerTransport}. Here, we show that conducting fixed points exist not only for gapless but, in a sense, also for gapped spin chains. However, beyond a low-bias region with nearly ideal conductance the current-voltage curves at these fixed points are qualitatively different in the two regimes, with a smaller conductance in the gapped phase. 

The rest of this paper is organized as follows. 
In Sec.~\ref{sec2}, we introduce the model and describe the numerical method employed. 
We then demonstrate in Sec.~\ref{sec4} the existence of non-trivial conducting fixed points. 
 To this end, we calculate steady-state spin currents and Friedel oscillations at the interface. 
In Sec.~\ref{sec5}, current-voltage curves for the gapless and the gapped regime are examined. 
Finally, Sec.~\ref{sec6} summarizes our main results.

\section{Model and method}
\label{sec2}
We consider the spin transport through a spin-chain material 
sandwiched between two conducting leads. The transport is assumed to occur in the spin chain direction and all inter-chain couplings are neglected. 
Thereby, we end up with a one-dimensional Hamiltonian  
\begin{equation}
\hat{H}_0 = \hat{H}_{\rm S} + \hat{H}_{{\rm L}_1} + \hat{H}_{{\rm L}_2} + \hat{H}_{{\rm S-L}_1} + \hat{H}_{{\rm S-L}_2} \, ,
 \label{eqham}
 \end{equation}
with $\hat{H}_{\rm S}$ describing a single spin chain, $\hat{H}_{{\rm L}_1 ({\rm L}_2)}$ the left (right) lead, and $\hat{H}_{{\rm S-L}_1 ({\rm S-L}_2)}$ the coupling between the spin chain and the left (right) lead. 
From now on, we restrict ourselves to the spin-1/2 XXZ case so that
\begin{equation}
\hat{H}_{\rm S} =  J \sum_{j=1}^{N_{\rm S}-1} \left[ \frac{1}{2} \left( \hat{S}_j^+ \hat{S}_{j+1}^- + \hat{S}_j^- \hat{S}_{j+1}^+ \right) + \Delta \hat{S}_j^z \hat{S}_{j+1}^z \right] \, ,
\end{equation}
where $N_{\rm S}$ is the number of sites of the spin chain, $\hat{S}_j^\alpha$ is the $\alpha\,(=x,y,z)$ component of the spin-1/2 operator at site $j$, and $\hat{S}_j^\pm=\hat{S}_j^x\pm i \hat{S}_j^y$. 
The fermionic leads are modeled by semi-infinite tight-binding chains  
at half-filling. Thus, the Hamiltonian for the left (right) lead is
\begin{equation}
 \hat{H}_{{\rm L}_1({\rm L}_2)} = -t  \sum_{\sigma = \uparrow, \downarrow} \sum_{\begin{array}{c}\scriptstyle j <  0 \\[-1mm] \scriptstyle (j  >  N_{\rm S}) \end{array} }
 \left[ \hat{c}_{j\sigma}^{\dagger} \hat{c}_{j+1,\sigma}^{\phantom{\dagger} } + \hat{c}_{j+1,\sigma}^{\dagger} \hat{c}_{j\sigma}^{\phantom{\dagger} } \right] \, , 
\end{equation}
where $ \hat{c}_{j\sigma}$ is the annihilation operator of an electron at site $j$ with spin $\sigma\,(=\uparrow,\downarrow)$. 
For simplicity, the couplings between the spin chain and the leads are assumed to be identical to the exchange interaction inside the spin chain. 
By defining the spin operators $\hat{S}_j^+=\hat{c}_{j\uparrow}^{\dagger}\hat{c}_{j\downarrow}^{\phantom{\dagger}}$, $\hat{S}_j^-=\hat{c}_{j\downarrow}^{\dagger}\hat{c}_{j\uparrow}^{\phantom{\dagger}}$, and $\hat{S}_j^z = \frac{1}{2} \left( \hat{c}_{j\uparrow}^{\dagger}\hat{c}_{j\uparrow}^{\phantom{\dagger}} - \hat{c}_{j\downarrow}^{\dagger}\hat{c}_{j\downarrow}^{\phantom{\dagger}}  \right)$ at tight-binding site $j$, the coupling terms can be written as
\begin{align}
& \hat{H}_{{\rm S-L}_1} = J \left[ \frac{1}{2} \left( \hat{S}_{0}^+ \hat{S}_{1}^- + \hat{S}_{0}^- \hat{S}_{1}^+ \right) + \Delta  \hat{S}_{0}^z\hat{S}_{1}^z \right],  
\end{align}
and
\begin{align}
& \hat{H}_{{\rm S-L}_2} = J \left[ \frac{1}{2}  \left( \hat{S}_{N_{\rm S}}^+ \hat{S}_{N_{\rm S}+1}^- + \hat{S}_{N_{\rm S}}^- \hat{S}_{N_{\rm S}+1}^+ \right) + \Delta  \hat{S}_{N_{\rm S}}^z\hat{S}_{N_{\rm S}+1}^z \right] \, .
\end{align}

\begin{figure}[bt]
\centering
\includegraphics[width=0.82\linewidth]{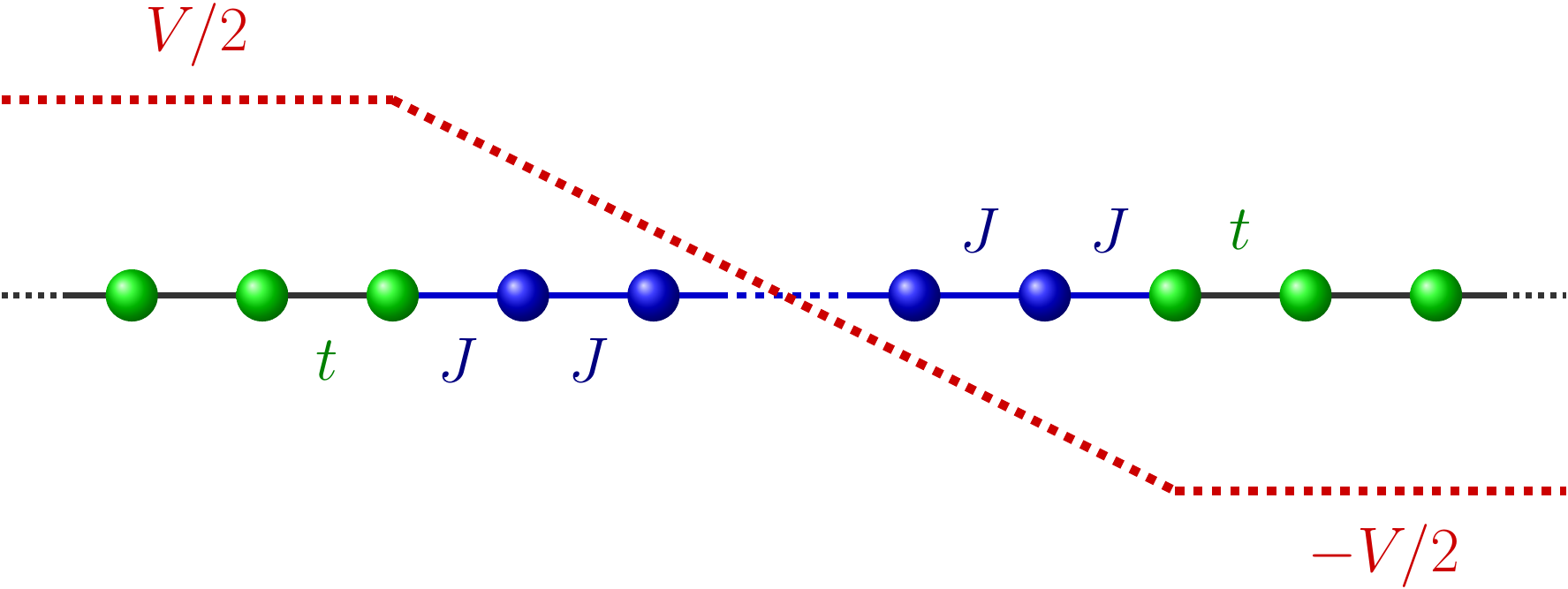}
\caption{Schematic depiction of the setup defined by the Hamiltonian ${\hat H}_0+{\hat H}_{\rm V}$ 
according to Eqs.~\eqref{eqham} and \eqref{eqpert}. Blue (green) circles indicate the spin chain 
(left and right leads). The red dashed line denotes the spin bias potential, which linearly decreases 
inside the spin chain. 
}
\label{fig1}
\end{figure}

We calculate the steady-state spin current that is generated by applying a spin bias voltage $V$.  
As in Ref.~\onlinecite{MeisnerTransport}, it is assumed that the potential drops off linearly in the spin chain, 
which adds the following term in the Hamiltonian (see also Fig.\ref{fig1}): 
\begin{equation}
\hat{H}_{\rm V} = \sum_{j} V_j \hat{S}_j^z , 
\label{eqpert}
\end{equation}
where
\begin{equation}
 V_j = \begin{cases} \frac{V}{2} \,, &  j \le0 \\[1mm]  
-\frac{V}{N_{\rm S}+1}j + \frac{V}{2} \,, & 1 \leq j \leq N_{\rm S} \\[1mm]
-\frac{V}{2} \,, &  j\ge  N_{\rm S}+1 \, . \end{cases} 
\end{equation}
The operator of the local spin-current is defined as
\begin{gather}
\hat{j}_j^{z} = \begin{cases} 
- \dfrac{it}{2} \hat{\bm{c}}_j^{\dagger} \sigma_z \hat{\bm{c}}_{j+1}^{\phantom{\dagger}} + {\rm h.c.} \,, & j < 0 \ {\rm  or }\  j > N_{\rm S} \\[3mm] \phantom{-} \dfrac{iJ}{2}\hat{S}_j^+ \hat{S}_{j+1}^- + {\rm h.c.}\, , &  0 \leq j \leq N_{\rm S} \, , \end{cases} 
\end{gather}
where $\hat{\bm{c}}_j^{\dagger} = ( \hat{c}_{j\uparrow}^{\dagger} ,\hat{c}_{j\downarrow}^{\dagger} )$ and 
$\sigma_z$ is the $z$ component of Pauli matrices~\cite{CurrentOperatorHubbard}. 
Our transport simulations are carried out in the zero-temperature limit. Then the system is initially in the ground state at time $\tau=0$. 
More precisely, the time evolution is started from the ground state of $\hat{H}_0$, 
where the spin chain and the leads are already coupled, and the spin bias voltage $V$ is applied at $\tau=0$.  
As discussed in Refs.~\onlinecite{ReviewSchmitteckert,JeckelmannTransport}, other setups are possible. 
For example, 
if one starts with the two leads decoupled from the spin chain and turns on the coupling, 
the transient behavior is different but the same steady-state properties are obtained. 
If, instead, the system is in the ground state with a finite spin bias $V$ and the bias is switched off at $\tau=0$, different steady-state currents are expected for large $V$~\cite{JeckelmannTransport}.

For the numerical calculation of the steady-state current, we mostly follow the MPS-based approach of Refs.~\onlinecite{ReviewSchmitteckert,MeisnerTransport,JeckelmannTransport}. 
The DMRG and parallel TEBD are used, respectively, to calculate the ground state of ${\hat H}_0$ 
and simulate the time-evolution after the spin bias (described by ${\hat H}_{\rm V}$) is switched on at $\tau=0$. 
We employ a standard Suzuki-Trotter approximation where the Hamiltonian is decomposed into terms acting on even and odd bonds. Specifically, a second-order decomposition with time step $\delta \tau  = 0.05/t$ is used. 
The leads have to be truncated to finite length $N_{\rm L}$, which gives rise to a discretization in the energy spectrum. 
The error due to this may be reduced by choosing appropriate boundary conditions with bond-dependent hopping 
strength that increase the energy resolution in the relevant energy region~\cite{DMRGKubo,ReviewSchmitteckert}.
Here, however, we find the leads with uniform hopping $t$ to be sufficient.

In our calculation of the steady-state current, the accuracy is mainly limited by the accessible time scale. 
The finite size of the leads obviously restricts the simulations to the time until the current reflected at the open 
boundaries of the leads returns to the spin chain. 
Additionally, the entanglement growth of an out-of-equilibrium state requires an increase of the bond dimension 
$m$ during the course of the time evolution, which eventually makes an accurate MPS representation of the state 
too costly. 
In the current setup, the von-Neumann entanglement entropy of the state after the perturbation grows linearly 
with the time~\cite{MeisnerTransport}, which  requires an exponential increase of the bond dimension $m$ for a 
fixed truncation error. 
The rate of the entanglement growth depends strongly on the applied voltage $V$. 
Simulation for large $V$ are typically more expensive. 
We fix the truncation error to a maximum discarded weight $10^{-6}$, which, in the worst cases, 
requires bond dimensions as large as $m=2200$.

In principle, an MPS representation with one tensor for each site $j$ in Eq.~\eqref{eqham} could be used for all of 
our simulations. 
However, for small $V$, where larger lead sizes are necessary to get accurate results, 
we find it advantageous to split the tight-binding leads into two branches with different $z$ component of the spin 
and employ a tree-tensor-network description~\cite{TTNGeneral} analogous to Ref.~\onlinecite{TTNImpurity}. 
This algorithm scales as $m^4$ at the interfaces, 
instead of $m^3$, but the representation of the 
tight-binding leads becomes much more efficient, allowing us to simulate larger leads. 
In addition, the worse scaling of the bond dimension $m$ is softened by the fact that the entanglement entropy 
at equilibrium is smallest at the interfaces, as already observed in Ref.~\onlinecite{MeisnerTransport}.

\section{Conducting fixed point}
\label{sec4}
Both the tight-binding chain and the spin-1/2 XXZ chain in the gapless regime are ballistic spin conductors at zero temperature. 
However, when these systems form a junction as described in Eq.~\eqref{eqham}, the transport may be supressed by scattering at the interfaces. 
For different, purely fermionic junctions 
a field-theoretical analysis has shown that the relevant backscattering that leads to insulating behavior at low temperatures vanishes for certain values of the model parameters~\cite{ConductingFixedPoint1,ConductingFixedPoint2,ConductingFixedPoint3}. 
At these conducting fixed points, the effective low-energy field theory is an inhomogeneous Luttinger liquid (LL). 
One may expect to find similar conducting fixed points for the spin-chain junction, since the gapless XXZ chain and the spin sector of the tight-binding leads are separately described by Luttinger liquids~\cite{GiamarchiBook}. In this section, we numerically show that such conducting fixed points indeed exist. 
The LL description of our model is given in Appendix~\ref{appA}. A proper field-theoretical treatment of the junction between spin-chain and tight-binding lead is left for a future investigation. 

\subsection{Spin current}

\begin{figure}[bt]
\centering
\includegraphics[width=0.9\linewidth]{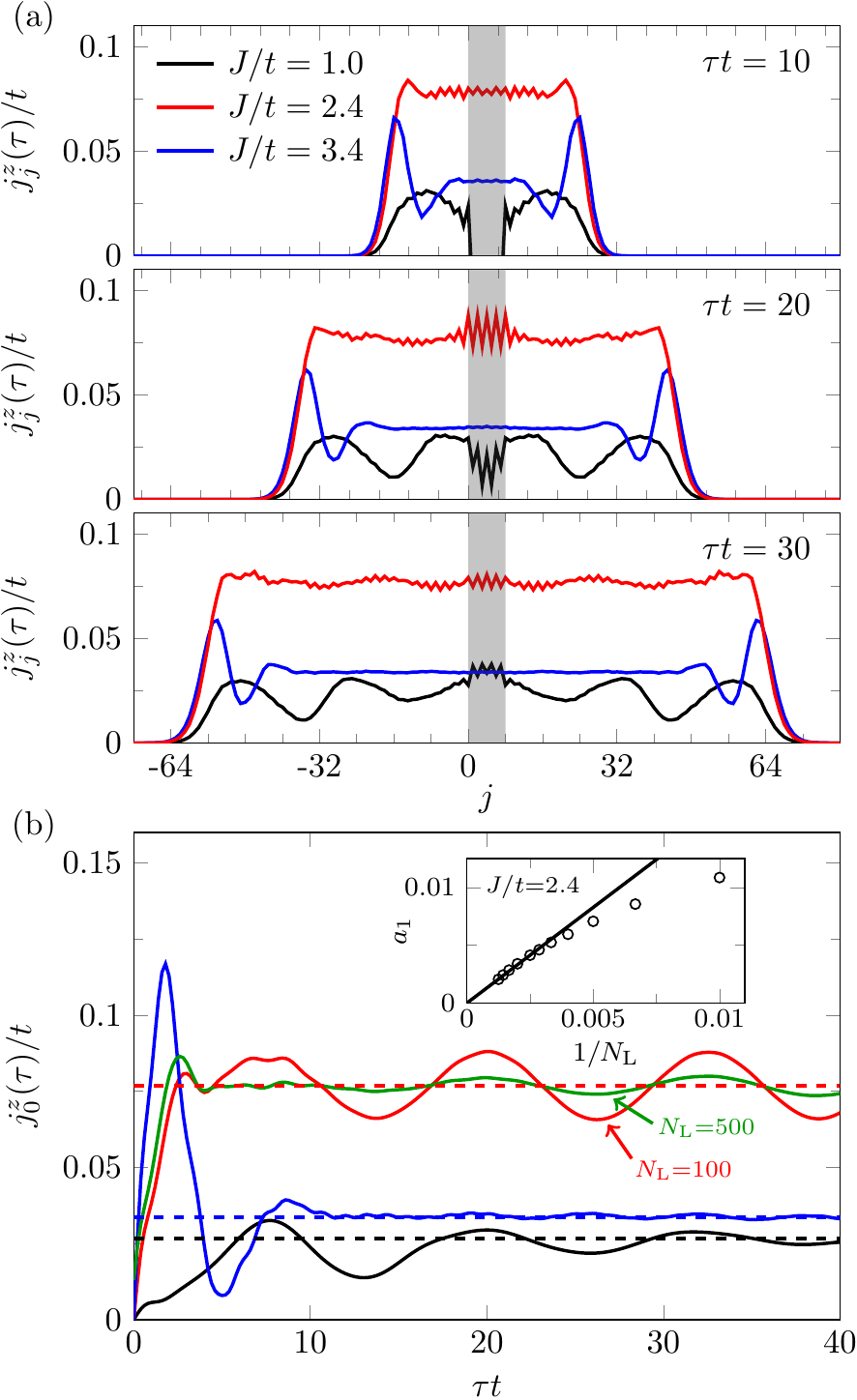}
\caption{Time evolution of the spin current $ j_j^z(\tau)$ in a junction composed of an isotropic spin chain ($\Delta=1$) of $N_{\rm S}=8$ sites (shaded region) coupled to tight-binding leads of $N_{\rm L}=100$ sites for spin bias $V/t=1$ and several values of $J/t$.  
(a) Spin current profile at three different times $\tau t = 10$, 20, and 30. 
(b) Time dependence of the spin current $j_0^z(\tau)$ between the spin chain and the left lead (solid lines) and estimated steady-state value (dashed lines). 
The result for $J/t=2.4$ with a larger size of the leads $N_{\rm L}=500$ is indicated by the green line. 
In the inset, the amplitude $a_1$ of the current oscillations [see Eq.~\eqref{eqfit}] is shown for several different lead sizes $N_{\rm L}$. The solid line is a fit to $a_1 \propto 1/N_{\rm L}$. }
\label{fig2}
\end{figure}

To search for conducting fixed points, we simulate the spin transport at finite spin bias 
for the two-lead setup described in Eq.~\eqref{eqham}. 
Let us first illustrate the procedure used to obtain the steady-state spin current. 
Figure~\ref{fig2}(a) shows the spin current profile for different time $\tau$ after the spin bias is switched on at $\tau=0$. 
The current starts to flow in the spin chain and spreads over to the leads, where the wavefront moves with the Fermi velocity 
$2t$. While the spin current in the spin chain becomes position and time independent in a true steady state, 
we find it fluctuating even at the maximum simulated time. 
Therefore, here the steady-state value is 
estimated from the time dependence of the spin current $j_0^z(\tau)$ between the spin chain and the lead, 
as demonstrated in Fig.~\ref{fig2}(b) for the isotropic chain. 
After a transient time of $\tau t \approx 10$, the spin current oscillates around its steady-state value 
with a period of approximately $4\pi/V$. 
This kind of oscillation has been explained as a Josephson current that arises because of the finite size of the leads 
and the corresponding gap between the single-particle energy levels~\cite{ReviewSchmitteckert}. 
We calculate the steady-state value of the spin current either by simply averaging $j_0^z(\tau)$ over multiple periods 
of the oscillation or by adapting it to
\begin{equation}
 j_0^z(\tau) = a_0 + a_1 \cos(\tau  V / 2 + a_2) \,,
\label{eqfit}
\end{equation}
where $a_0,a_1$ and $a_2$ are fit parameters~\cite{ReviewSchmitteckert}.

The spin current generally depends on both the anisotropy $\Delta$ and the ratio $J/t$ of the exchange 
interaction in the spin chain and the hopping amplitude in the leads. 
For most of the parameter space, the spin current is expected to be strongly suppressed 
because of the backscattering at the interfaces. 
As we will show, however, the system can be tuned to a conducting fixed point for each $\Delta$ by varying $J/t$. 
In the isotropic chain $(\Delta=1)$ considered in Fig.~\ref{fig2}, for example, the corresponding value is 
$(J/t)_{\rm c} \approx 2.4$. 
The current there is much larger than for the other values shown, $J/t=1$ and $J/t=3.4$, which lie away from the conducting fixed point. 

The ratio $J/t$ affects not only the steady-state value of the spin current but also the oscillation of the current 
as a function of time $\tau$. 
For a fixed size of the leads with $N_{\rm L}=100$, the current oscillation at the interface is strongest 
at $J/t = (J/t)_{\rm c}$ where  
it appears nearly undamped [see Fig.~\ref{fig2}(b)]. 
For either larger or smaller value of $J/t$, on the other hand, the oscillation decays relatively quickly with increasing $\tau$. 
By using the tree-tensor-network method, we also consider a junction with much larger leads of $N_{\rm L}=500$ sites. 
In this case, the current oscillation at the conducting fixed point becomes significantly smaller, 
as shown in Fig.~\ref{fig2}(b), which confirms that it is mostly caused by the discretization of 
the single-particle energy levels in the leads. 
It is expected that the amplitude of the oscillations is proportional to the gap between single particle levels and thereby inverse proportional to $N_{\rm L}$~\cite{ReviewSchmitteckert}. 
As shown in the inset of Fig.~\ref{fig2}, 
this agrees with our results for $N_{\rm L} \gtrsim 400$, while deviations are seen for smaller leads. 
The steady-state values of the spin current estimated from the simulations are the same for each $N_{\rm L}$.

\begin{figure}[tb]
\centering
\includegraphics[width=0.95\linewidth]{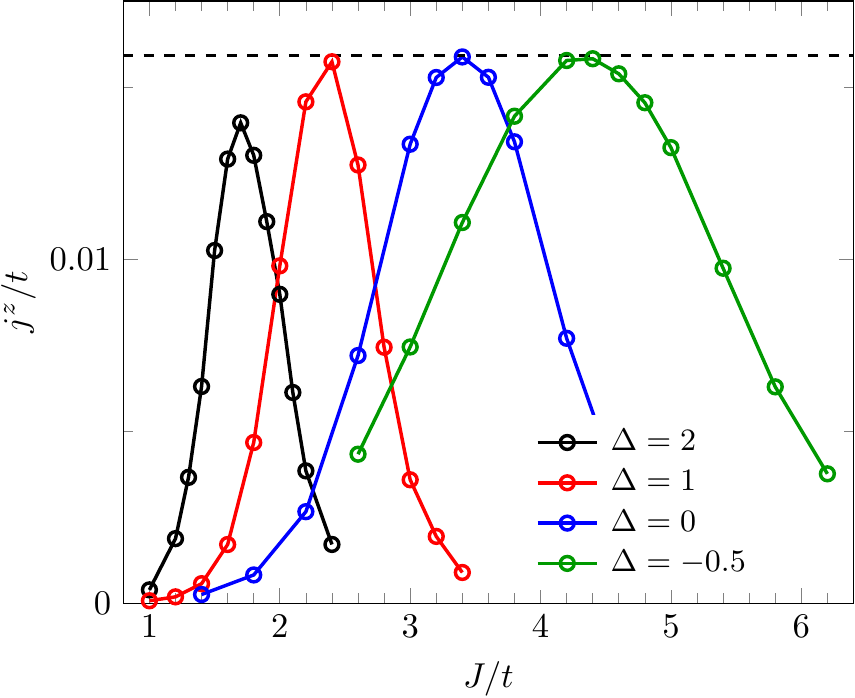}
\caption{
Steady-state spin current $j^z$ as a function of $J/t$ for four different values of $\Delta$. Other parameters are $N_{\rm S}=8$, $N_{\rm L}=500$, and $V/t=0.2$. 
The dashed line shows the current $V/(4\pi)$ expected for a Luttinger liquid with 
smooth interfaces. 
}
\label{fig3}
\end{figure}

Figure~\ref{fig3} shows the dependence of the steady-state spin current $j^z$ on the ratio $J/t$ at the 
fixed spin bias voltage $V/t=0.2$ for several values of $\Delta$. 
In each case, a clear maximum of the spin current appears. 
We first address the gapless phase for $\Delta = 1$, 0, and $-0.5$ where the LL description is applicable. 
For parameters in this regime, the maximum current obtained is close to $V/(4\pi)$, which, as discussed in 
Appendix~\ref{appA}, is the current for a LL with adiabatic contacts. 
This indicates that a conducting fixed point with ideal linear conductance exists at the ratio $(J/t)_{\rm c}$ which maximizes the current. 
As $\Delta$ is decreased, $(J/t)_{\rm c}$ becomes larger. 
In addition, the current peak as a function of $J/t$ broadens, 
which suggests that the backscattering becomes less relevant. 
A current maximum remains, however, even for negative $\Delta$.

Figure~\ref{fig3} also shows the results for $\Delta=2$ in the gapped phase. 
While a sharp peak is still observed, the maximum value of the spin current does not reach the ideal value in this case. 
The vanishing of the Friedel oscillations (see Sec.~\ref{secFriedel}) for the parameters at the current peak indicates that the relevant backscattering at the interfaces can still be tuned to zero. 
Therefore, the deviation from the ideal conductance appears to be caused by 
different reasons, most likely related to properties in the bulk of the spin chain, which for $\Delta > 1$ is no longer described by a LL model. 
How the spin transport differs in the gapped and gapless phases of the antiferromagnetic XXZ chain will be 
analyzed in Sec.~\ref{sec5}.

\subsection{Friedel oscillations}
\label{secFriedel}

Besides its effect on the transport, the backscattering at inhomogeneities 
is known to induce characteristic Friedel oscillations of the local density or magnetization 
with twice the Fermi wavenumber $k_{\rm F}$~\cite{RommerEggert2000}. 
The Friedel oscillations at the interface vanish, however, if the backscattering amplitude is tuned to zero. 
The calculation of the magnetization profile therefore constitutes a different, perhaps more efficient way 
to search a conducting fixed point~\cite{ConductingFixedPoint1}. 
As a consistency check for the results of the spin-transport simulations above, 
we now investigate the dependence of the Friedel oscillations on $J/t$ for fixed $\Delta$ with no spin bias applied. 
Since the magnetization is uniform in the spin-flip symmetric case, we examine 
the local susceptibility~\cite{ConductingFixedPoint1} instead by adding a small uniform magnetic field 
described by $\delta \hat{H} = h \sum_j \hat{S}_j^z$. 
For these calculations, we consider a single interface between the tight-binding lead and the spin chain 
because the Friedel oscillations typically decay over a distance longer than the spin-chain length 
accessible in our transport simulations. Furthermore, we consider finite temperatures by using the grand-canonical 
purification method~\cite{MPSPurification}, which avoids problems in the convergence of the DMRG 
ground-state calculations. 
The purification method allows us to keep track of the growth of the Friedel oscillations starting from the interface and the open 
ends of the system as the temperature is lowered successively. We terminate the simulations when the 
finite system size begins to affect the results. 
The finite-temperature calculations also allow us to study the gapped phase of the spin chain where the ground state 
is antiferromagnetically long-range ordered.

\begin{figure}[tb]
\centering
\includegraphics[width=0.99\linewidth]{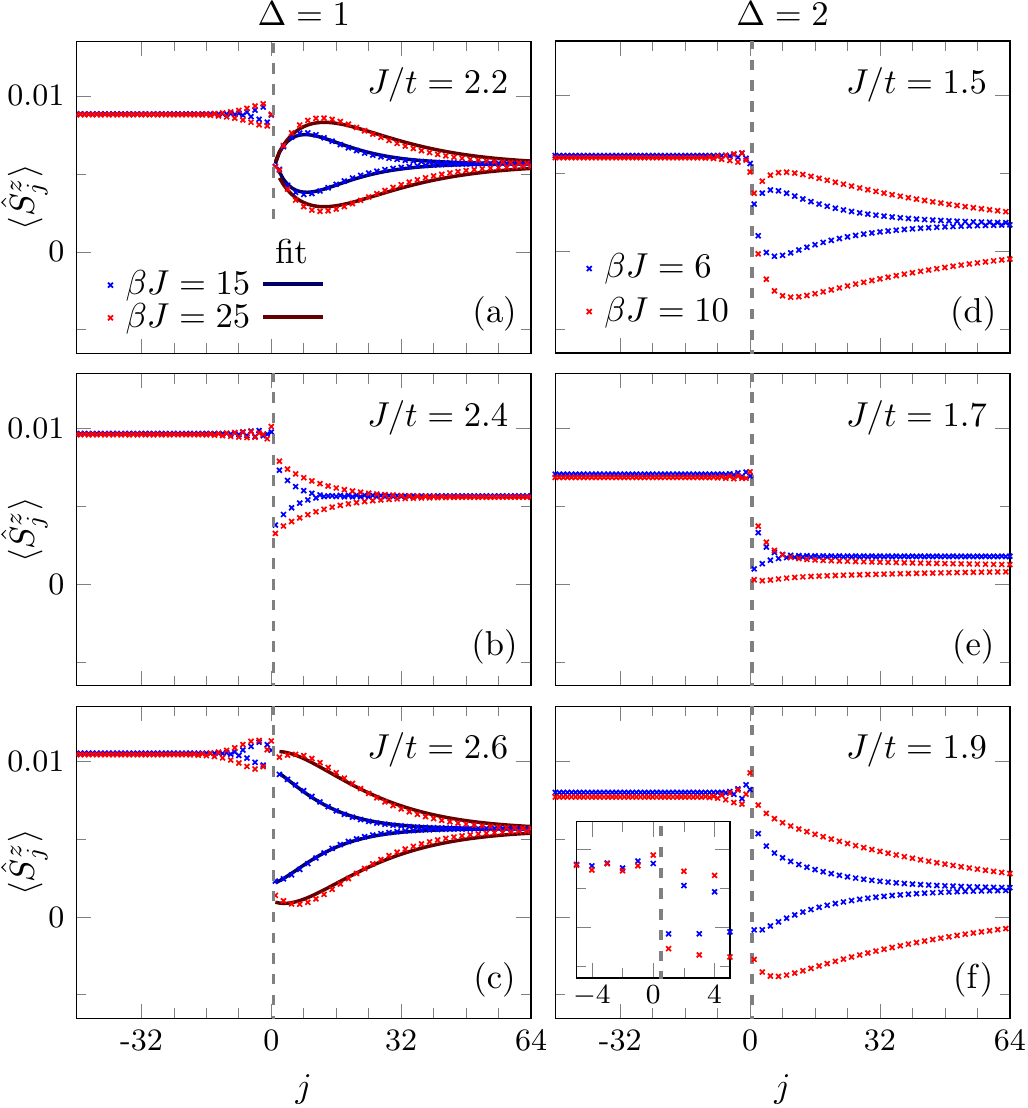}
\caption{Magnetization profile $\langle {\hat S}^z_j\rangle$ around the interface for an applied magnetic field $h/J=0.05$. 
The dashed line indicates 
the interface between the tight-binding lead $(j \leq 0)$ and the spin chain $(j > 0)$. The systems sizes are 
$N_{\rm L}=400$ and $N_{\rm S}=400$ for $\Delta=1$ and $N_{\rm L}=400$ and $N_{\rm S}=800$ for $\Delta=2$. 
The inset in (f) is a magnified view of the region close to the interface, highlighting the Friedel oscillations 
with wavenumber $\pi$. 
Solid black lines in (a) and (c) are fits of the data in the form of Eq.~\eqref{eqfitFriedel}. 
The results are obtained by finite temperature calculations at the inverse temperature 
$\beta$. 
}
\label{fig4}
\end{figure}

Figure~\ref{fig4} shows the magnetization profile around the interface for the magnetic field strength $h/J=0.05$.
Here, we fix $h/J$ instead of $h/t$ because for the values of the anisotropy $\Delta$ considered, 
the Friedel oscillations are much stronger in the spin chain than in the lead. 
Since the spin chain without magnetic field corresponds to a half-filled chain of fermions, 
the local magnetization oscillates with wavenumber $2k_{\rm F}=\pi$. 
As expected, the effect is larger at low temperatures. 
For the fixed exchange anisotropy, the strength of the Friedel oscillations has a minimum as a function of $J/t$. 
This behavior can be observed in both the gapless and gapped regimes. 
For the former case, we have attempted a fit to the oscillation profile 
\begin{align}
\chi(j+\tilde{a}_0) = \tilde{a}_1 T^{\bar{K}-1} j (-1)^j & \left[\frac{v}{T}  \sinh  \left(\frac{2 \pi T j}{v} \right) \right]^{-K} \nonumber \\
&\times P_{-\bar{K}}(\coth(2\pi T j / v))
\label{eqfitFriedel}
\end{align}
derived for the susceptibility of a chain of spinless fermions with an abrupt jump of the parameters~\cite{ConductingFixedPoint1,ConductingFixedPoint2}. 
Here, $P_l(z)$ is the Legendre function, $K$ and $v$ are the LL parameter and the spin velocity, respectively, and $\bar{K}$ is determined by the LL parameters on both sides of the interface (see Appendix~\ref{appA}). Free parameters of the fit are a position offset $\tilde{a}_0$ and the amplitude $\tilde{a}_1$. The fits for the even and odd sites separately are shown in Figs.~\ref{fig4}(a) and \ref{fig4}(c)
for the oscillations in the spin chain with $\Delta = 1$, where $K=1/2$ and $v=J \pi/2$, and we 
set $\bar{K}=1/2$, corresponding to an isotropic spin chain with a jump in the exchange parameter. 
Very good agreement is found with our numerical data, suggesting that Eq.~\eqref{eqfitFriedel} or a similar relation is also applicable to the junction with the fermionic lead.

To measure the overall strength of the Friedel oscillations, we introduce a quantity
\begin{equation}
O_{\rm F} = \sum_{j=1}^{N'} |\langle \hat{S}_{j+1}^z - \hat{S}_{j}^z \rangle | \, ,
\label{eqOF}
\end{equation}
where $N'$ is chosen so that the Friedel oscillations due to the open boundary at the end of the spin chain are excluded. 
By calculating $O_{\rm F}$, we search for a value of $J/t$ that minimizes the Friedel oscillations for a fixed anisotropy 
$\Delta$.

\begin{figure}[bt]
\centering
\includegraphics[width=0.9\linewidth]{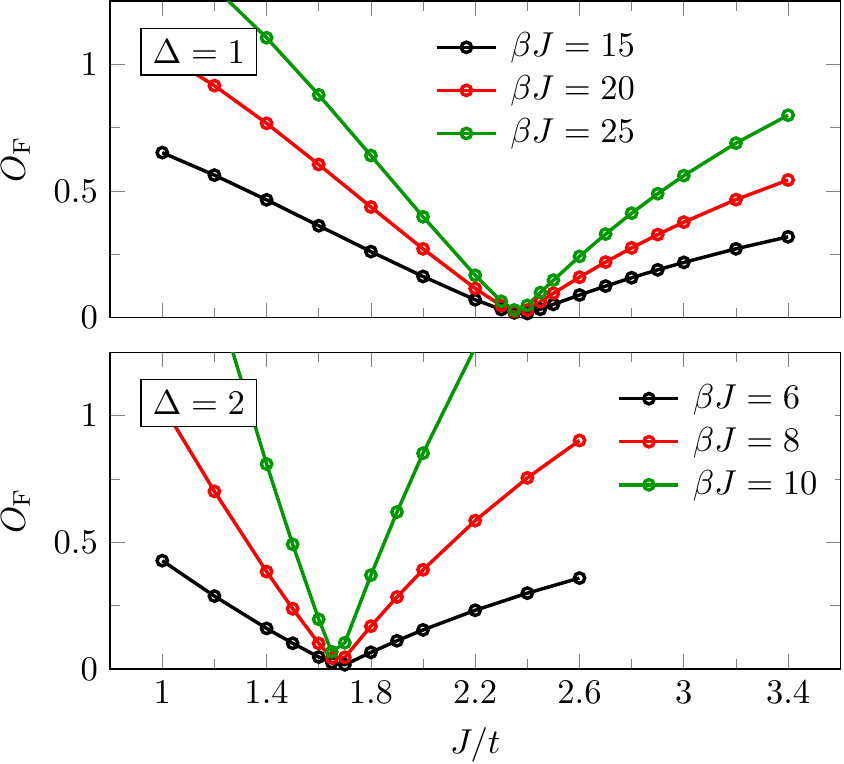}
\caption{Strength of the Friedel oscillations, $O_{\rm F}$, defined in Eq.~\eqref{eqOF} around the interface 
inside the spin chain at the inverse temperature $\beta$. The system sizes are the same as in Fig.~\ref{fig4}. 
}
\label{fig5}
\end{figure}

The results for $\Delta=1$ and $\Delta=2$ are shown in Fig.~\ref{fig5}. 
In all cases studied, including the gapped regime, we find a clear minimum of the Friedel oscillation strength. 
where approximately $O_{\rm F}=0$, which suggests that the relevant backscattering vanishes. 
When the temperature is lowered, the position of the minimum moves to smaller $J/t$. 
The temperature dependence seems to be stronger for small $\Delta$. 
By identifying the position of the minimum for $T \rightarrow 0$ as the conducting fixed point, 
we obtain $(J/t)_{\rm c}\approx 2.4$ for $\Delta=1$. 
This value agrees with the results of the spin-transport simulations for $N_{\rm S}=8$, despite the fact 
that we now consider the limit of a large spin chain. 
Identifying $(J/t)_{\rm c}$ similarly in the gapped phase, we obtain $(J/t)_{\rm c}\approx 1.7$ 
for $\Delta=2$, which also coincides with the value of $J/t$ where the spin current becomes maximum in Fig.~\ref{fig3}. 
When calculating $(J/t)_{\rm c}$ as a function of the anisotropy $\Delta$, we find no qualitative difference 
across the phase boundary at $\Delta=1$.

\section{Current-voltage characteristics}
\label{sec5}

Having established the existence of conducting fixed points with a finite linear conductance in the previous section, 
we now turn our attention to the spin-bias dependence of the spin current. 
To examine how the current-voltage curve is modified by the backscattering at the interfaces and the presence of 
a finite energy gap, 
the system parameters at and away from the line of conducting fixed points are considered for both the gapless and 
gapped phases of the antiferromagnetic spin-1/2 XXZ chain. 

\subsection{Gapless regime} 

\begin{figure}[bt]
\centering
\includegraphics[width=0.92\linewidth]{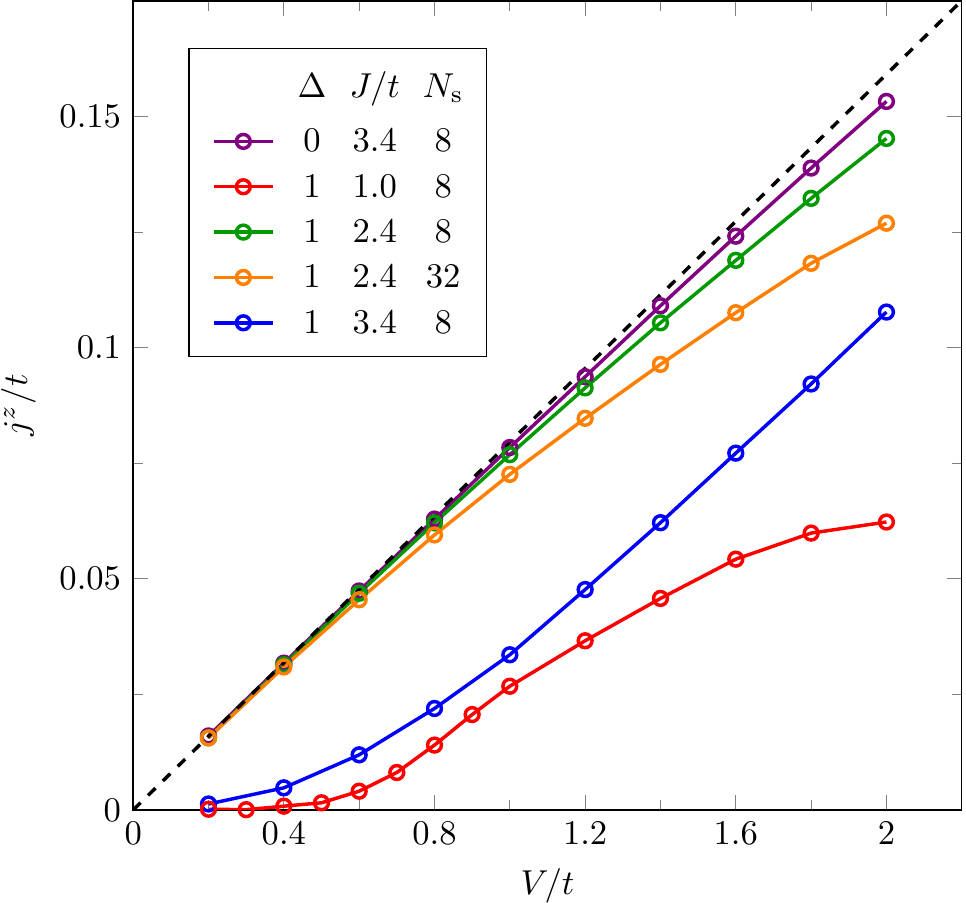}
\caption{Current-voltage curve in the two-lead setup described in Eq.~\eqref{eqham} for different parameters 
in the gapless phase. The dashed line is the conductance $G=1/(4\pi)$ of a Luttinger liquid smoothly 
connected to non-interacting leads. }
\label{fig6}
\end{figure}

First, we study the gapless XY phase where the spin chain can be described by a LL model. 
As mentioned in Sec.~\ref{appA}, a spin conductance $G=1/(4\pi)$ is expected unless the transport is hindered by 
the backscattering at the interfaces. 
We have already confirmed that this ideal value can be obtained approximately at low spin bias $V/t=0.2$ 
by tuning $J/t$ to a conducting fixed point $(J/t)_{\rm c}$ for a given anisotropy $\Delta$. 
By calculating the current-voltage curve, we can determine at what energy scale the LL description becomes invalid and 
the linear behavior breaks down. 
Figure~\ref{fig6} shows the results for the isotropic spin chain $(\Delta = 1)$ and the XX spin chain $(\Delta=0)$ 
where the conducting fixed points are $(J/t)_{\rm c} \approx 2.4$ and $(J/t)_{\rm c} \approx 3.4$, respectively 
(see Fig.~\ref{fig3} and Fig.~\ref{fig5}).  
In both cases, the current-voltage curve for $J/t \approx (J/t)_{\rm c}$ shows good agreement with the LL prediction 
up to at least $V/t=1$, despite the strong inhomogeneity at the interfaces. 
For $\Delta=1$, increasing the length of the spin chain to $N_{\rm S}=32$ leads to stronger deviations at large $V$ while the currents for $V/t\lesssim 0.4$ remain nearly unchanged. 
Possible length-dependent corrections to the conductance have been considered, for example, in Refs.~\onlinecite{ConductanceDamping,PhysRevLett.107.056402}.

Away from the conducting fixed points, the low-bias conductance is strongly reduced by backscattering. 
This is demonstrated in Fig.~\ref{fig6} for an isotropic chain and values $J/t=1$ and $3.4$ that are significantly smaller 
or larger than $(J/t)_{\rm c} \approx 2.4$. 
In a LL with an impurity, the differential conductance eventually approaches the ideal value $1/(4\pi)$ with a power law 
as the bias in increased~\cite{Fisher1997}. 
This is consistent with our results for $J/t=3.4$ where an approximately linear current-voltage relation is restored 
for $V/t\gtrsim 1.2$.  
For $J/t=1$, on the other hand, the differential conductance drops off again at $V/t \approx 1$, 
likely because the bias voltage considered is already comparable or larger than the exchange constant $J$. 
In any case, the current should vanish in the large-$V$ limit for the chosen setup because of the finite bandwidth of the 
leads. 
This does not apply, however, to the setup where the spin voltage $V$ is present initially and then turned off at $\tau=0$~\cite{JeckelmannTransport,ReviewSchmitteckert}. 

\subsection{Gapped regime}

In Sec.~\ref{sec4}, it was shown that the finite-temperature Friedel oscillations around the interfaces can be tuned 
to zero by varying $J/t$ even in the gapped phase. 
Therefore, a fixed point $(J/t)_{\rm c}$ with vanishing relevant backscattering seems to exist in this regime as well. 
One may then ask how the current-voltage curve there differs from that at a conducting fixed point in the gapless phase. 
In the following, we examine this for the anisotropy parameter $\Delta=2$ where the Friedel oscillations disappear 
at $J/t \approx 1.7$. 

\begin{figure}[tb]
\centering
\includegraphics[width=0.92\linewidth]{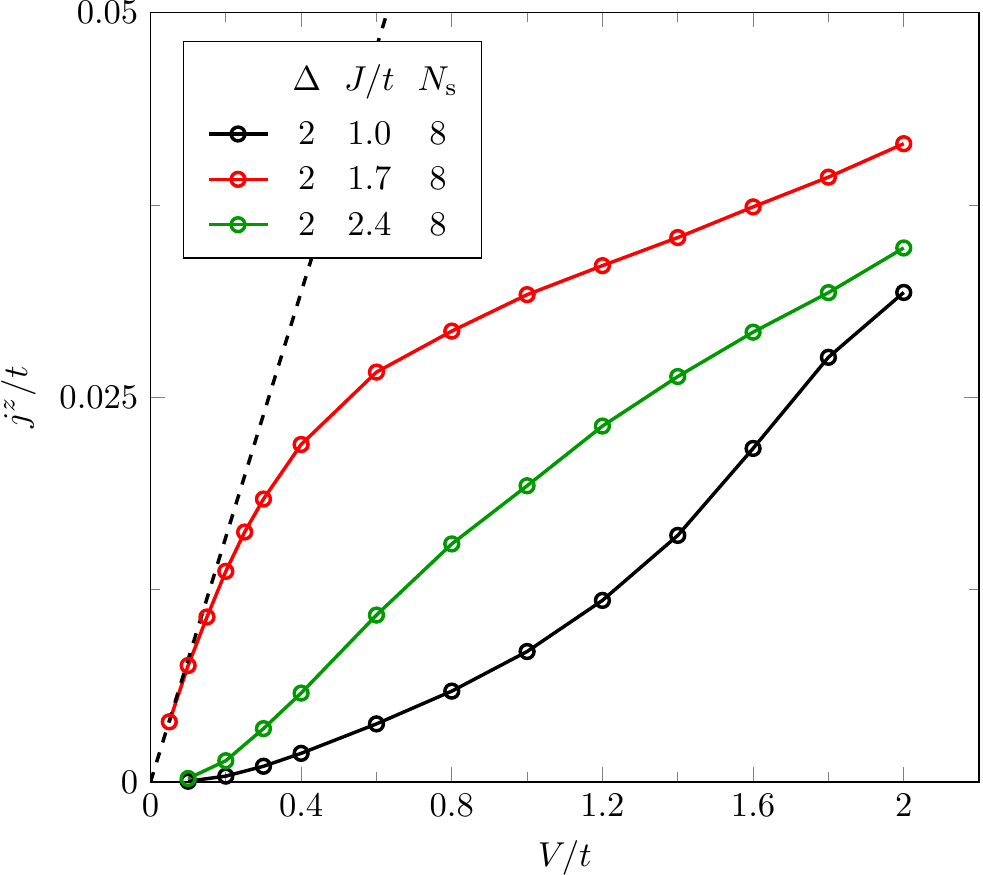}
\caption{Current-voltage curve for a spin chain with $N_{\rm S}=8$, anisotropy $\Delta=2$, and 
different values of $J/t$. The dashed line corresponds to the ideal conductance $G=1/(4\pi)$ obtained in the 
Luttinger-liquid regime. }
\label{fig7}
\end{figure}

\begin{figure}[t!]
\centering
\includegraphics[width=0.92\linewidth]{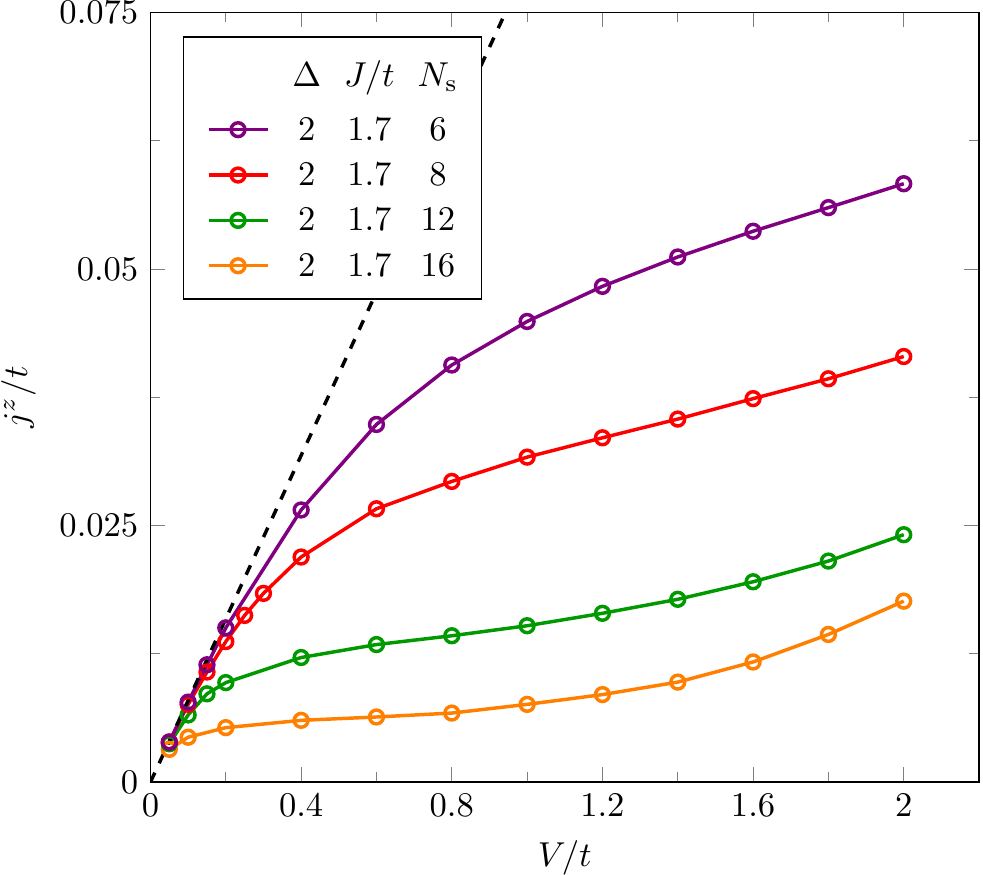}
\caption{
Same as in Fig.~\ref{fig7} but for a spin chain with $J/t=1.7$ and various chain lengths. 
}
\label{fig8}
\end{figure}

Figure~\ref{fig7} displays the current-voltage curve of a spin chain with $N_{\rm S}=8$ sites for $J/t = 1.7$ 
as well as for smaller and larger values of $J/t$. 
For $J/t = 1.7$, the conductance appears to approach $1/(4\pi)$ as the voltage $V$ is decreased to zero, 
indicating that almost ideal spin transport can be achieved at low energy. 
At larger voltage, on the other hand, the differential conductance drops off sharply, which is not observed 
in the LL phase. This crossover occurs approximately at $V/t\approx 0.4$. 
As in the LL regime, the spin current at small bias voltage is strongly reduced away from $(J/t)_{\rm c}$. 
Since the XXZ spin chain with $\Delta > 1$ is a spin insulator, 
the spin transport at fixed $V$ 
should become more and more suppressed with increasing the system size $N_{\rm S}$. 
The effect of $N_{\rm S}$ on the current-voltage curve for $J/t=1.7$ is shown in Fig.~\ref{fig8}. 
As expected, the spin current becomes noticeably smaller when going to larger system sizes $N_{\rm S}$. 
There is still a crossover below which the perfect spin conductance seems to be approached. 
However, this crossover is shifted to very small bias voltage $V$ with increasing $N_{\rm S}$.

Similar behavior, i.e, unsuppressed current for small systems at low energy, 
occurs in the charge transport through Hubbard chains with an odd number of sites~\cite{Oguri2001}. 
Perhaps more relevant to our model, such effect has been predicted for one-dimensional 
charge-density-wave insulators adiabatically contacted to non-interacting leads, by using field 
theoretical methods~\cite{Ponomarenko1999}. 
This model may be interpreted as a XXZ spin chain with the anisotropy $\Delta$ set to zero 
outside a finite region with $\Delta > 1$ that corresponds to the charge-density-wave part. 
In contrast to our results, a negative differential conductance was obtained. 
However, this may be related to the different choice of the leads.

For sufficiently long spin chains, we observe an upturn of the spin current at large spin bias. 
A setup analogous to ours has been considered in the calculation of the charge current through a Mott-insulating 
Hubbard chain connected to non-interacting leads~\cite{MeisnerTransport}. It was shown that the current-voltage 
curve can be described by a function 
$j(V) = a V e^{-V_{\rm c}/V} \, , $
where $a$ and $V_{\rm c}$ are constants. In particular, $V_{\rm c}$ is approximately proportional to the square of 
the charge gap of the disconnected Hubbard chain. This relation was previously obtained for the current in a periodic 
chain and explained in terms of a Landau-Zener mechanism~\cite{Oka}. 
The upturn observed for $N_{\rm S}=16$ in Fig.~\ref{fig8} suggests that a similar activated behavior 
occurs in our model for long enough chains where the low-voltage transport is suppressed. 
However, our available data is not sufficient to check the specific functional form and the dependence on the spin gap of the isolated spin chain.

\section{Conclusion}
\label{sec6}

We have numerically studied the finite-bias spin transport in a spin-1/2 XXZ chain connected to 
half-filled tight-binding leads at zero temperature, focusing on the effect of scattering at the interfaces. 
By calculating the steady-state spin current and the Friedel oscillations, 
it was shown that in the Luttinger liquid  regime, conducting fixed points with the ideal linear conductance exist, 
similarly as in related models for inhomogeneous quantum wires. 
Our results furthermore indicate that conducting fixed points also appear in the gapped phase. 
There, the nearly ideal spin transport can only be observed in a small bias voltage region, 
which shrinks when the length of the spin chain is increased. 

Our interpretation of the numerical data is partially based on the field-theoretical description which has been derived for a different type of junction consisting only of fermionic chains. 
 It would be interesting to find the effective low-energy field theory for the specific junction considered here, including explicit expressions for the scattering at the interfaces, 
and determine whether there are qualitative differences with the previously studied models. 

More difficult to treat numerically, but closer to actual 
experiments, is the finite temperature case. 
For the finite-temperature simulations, one could employ a similar TEBD method 
where the MPS describes a purification of the density matrix instead of a pure state. 
With the approach in Ref.~\onlinecite{KarraschThermal2013}, 
it may also be possible to study a setup where a spin current is driven by a temperature gradient, 
mimicking the experiment in Ref.~\onlinecite{SpinonSpinCurrent}.

In this paper, we have only considered junctions composed of spin-1/2 chains.
A possible extension would be to study analogous systems for spin ladders or chains with higher local spin. 
The spin-1 Heisenberg chain, for example, might be interesting since it is experimentally realizable and differs from the spin-1/2 chain in several aspects: Its elementary excitations are magnons instead of spinons, it is non-integrable, and it exhibits symmetry-protected edge states at open boundaries. 
For a setup with leads, the question then arises how the contact is affected by these edge states. This will addressed in a forthcoming study.

\section*{Acknowledgments}
DMRG simulations were performed using the ITensor library~\cite{ITensor}. 
F. L. was supported by Deutsche Forschungsgemeinschaft through project FE 398/8-1 and 
by the International Program Associate (IPA) program in RIKEN. This work was supported 
by Grant-in-Aid for Scientific Research from MEXT Japan under the Grant No. 17K05523 and also in part by RIKEN Center for Computational Science through the HPCI System Research projects hp120246, hp140130, and hp150140. 
T. S. acknowledges the Simons Foundation for funding.

\appendix

\section{Luttinger liquid description}
\label{appA}

The low-energy physics of the spin-1/2 XXZ chain in the XY phase $(-1 \leq \Delta \leq 1)$ are described by 
the Luttinger-liquid (LL) model~\cite{GiamarchiBook}
\begin{align}
\hat{H}_{\rm S} &= \frac{1}{2} \int dx \left[ \frac{v}{K}(\partial_x \phi)^2 + v K (\partial_x \theta)^2 \right] \, ,
\label{eqLL}
\end{align}
where the bosonic fields obey the commutation relations $[\phi(x),\partial_{x'}\theta(x')] = i\delta(x-x')$ 
and the LL parameter $K = \pi/[2 \arccos(-\Delta)]$ and the spin velocity 
$v = J \pi \sqrt{1-\Delta^2} /[2\arccos(\Delta)]$ are known from the Bethe-ansatz solution~\cite{LutherPeschel1975}. 
In this representation, the long-wavelength part of the magnetization is related to the fields by
\begin{equation}
\hat{S}^z(x) \simeq -\frac{1}{\sqrt{\pi}} \partial_x \phi \, .
\label{eqSpinOpLL}
\end{equation}
The charge transport in a system of spinless fermions with a nearest-neighbor interaction corresponds directly to 
the spin transport in the spin-1/2 XXZ chain since the models are related by a Jordan-Wigner transformation. 
For an infinite homogeneous chain, the spin conductance $G$ is given by~\cite{KaneFisher2} 
\begin{equation}
G=\frac{K}{2\pi} \, . 
\label{eqLLconductance}
\end{equation}
In general, however, this expression is no longer valid when leads are taken into account. 
The effective low-energy Hamiltonian of the tight-binding leads in our setup described by Eq.~\eqref{eqham} 
consists of two components of the form of Eq.~\eqref{eqLL} for the charge and spin sectors. 
Requiring the representation of the leads to be consistent with Eqs.~\eqref{eqSpinOpLL} and \eqref{eqLLconductance} fixes the spin LL parameter to $K=1/2$. This is also the value for the spin chain at the $SU(2)$ symmetric point $\Delta = 1$. 

A single junction between spin chain and lead has some similarity with the single-channel Kondo model, except that the impurity site is now also coupled to a spin chain.  
We assume that, analogously to the Kondo model, the charge and spin sectors are decoupled in the low-energy theory~\cite{AFFLECK1990517}. 
Focusing only on the spin part and ignoring any possible boundary terms, the naive field-theoretical description of our system becomes an 
inhomogeneous LL with the position-dependent LL parameter $K(x)$ and spin velocity $v(x)$. 
It has been shown that the conductance of such a system is obtained by replacing the LL parameter in 
Eq.~\eqref{eqLLconductance} with its asymptotic value in the leads $K(x\rightarrow \pm \infty)$~\cite{MaslovStone,SafiSchulz}. 
For the non-interacting leads, the spin conductance therefore is $G=1/(4\pi)$, independent of the parameters 
in the spin chain.

By using an inhomogeneous LL model to describe a one-dimensional junction
one assumes that backscattering at the interfaces can be neglected. 
This is justified for adiabatic contacts but not for the abrupt transition between the spin chain and the lead 
described in Eq.~\eqref{eqham}. 
For a chain of spinless fermions with uniform LL parameter $K$, the effect of backscattering at an inhomogeneity on the linear conductance $G$ is well-known~\cite{KaneFisher1,KaneFisher2}: 
At zero temperature, $G$ vanishes if the interactions are repulsive (i.e., $K<1$), 
while $G$ is not reduced for attractive interactions (i.e., $K>1$). 
An abrupt change in the system parameters of a quantum wire has a similar impact on the conductance, 
as has been studied for both spinless~\cite{ConductingFixedPoint1,ConductingFixedPoint2} and 
spinful~\cite{ConductingFixedPoint3} fermions using bosonization and quantum Monte Carlo methods. 
In those cases, whether the transport is suppressed at low temperatures depends on the LL parameters on 
each side of the interface.  
For the spinless model, the zero-temperature conductance vanishes for $\bar{K} < 1$, where 
$\bar{K}=2(\frac{1}{K_1} + \frac{1}{K_2})^{-1}$, and $K_{1}$ and $K_2$ are the LL parameters on the left and right sides 
of the interface~\cite{ConductingFixedPoint1}. 
However, it was also shown that, even for abrupt junctions, 
conducting fixed points may be obtained by tuning certain system parameters such as the hopping and 
interaction strengths~\cite{ConductingFixedPoint1,ConductingFixedPoint3}.  
At these conducting fixed points, the amplitude of the relevant backscattering becomes zero and thus 
the ideal conductance determined by the LL parameters of the leads is recovered at zero temperature. 
Note that there is still irrelevant scattering at the interfaces, which can affect the conductance at finite temperatures. 

In the spin-chain junction described in Eq.~\eqref{eqham}, 
the couplings between the subsystems are different than in the previously studied fermionic models. 
Therefore, it is not clear that the field-theoretical results in the previous studies   
apply similarly in our system. However, we demonstrate in the main text for several values of $\Delta$ that conducting fixed points with ideal spin transport exist. 
Since these fixed points are obtained by varying a single model parameter, there appears to be only one relevant perturbation at the interfaces, similarly as in the purely fermionic chains. 
For $\Delta = 1$, this may be expected by noticing that the spin-chain junction corresponds to a strong-coupling limit of the 
inhomogeneous half-filled Hubbard chain for which conducting fixed points have been reported in Ref.~\onlinecite{ConductingFixedPoint3}. 
By analogy with the fermionic models, we refer to the relevant perturbation at the interfaces as ``backscattering".

\end{document}